\documentclass[a4paper,10pt]{IEEEtran}
\usepackage[nocompress]{cite}
\usepackage[utf8]{inputenc} 
\usepackage{graphicx}
\usepackage[tight,TABTOPCAP]{subfigure}

\usepackage{xspace}

\usepackage{pifont}

\usepackage{ifpdf} 

\usepackage{amsmath}
\usepackage{amsfonts}

\usepackage[noend]{algorithmic}
\usepackage{algorithm}

\usepackage[small]{caption}

\usepackage{listings}
\usepackage[pdfpagemode=none,bookmarks=false]{hyperref}

\usepackage{balance}

\usepackage{dcolumn}
\newcolumntype{d}{D{.}{.}{5.3}}
\newcolumntype{e}{D{.}{.}{4.3}}
\newcolumntype{f}{D{.}{.}{3.3}}
\newcolumntype{g}{D{.}{.}{3.1}}
\newcolumntype{h}{D{.}{.}{2.2}}
\newcolumntype{i}{D{.}{.}{4.1}}

\makeatletter
\renewcommand\fs@ruled{\def\@fs@cfont{\bfseries}\let\@fs@capt\floatc@ruled
  \def\@fs@pre{\hrule height.8pt depth0pt \kern2pt}%
  \def\@fs@post{\kern2pt\hrule\relax\vspace{-5mm}}%
  \def\@fs@mid{\kern2pt\hrule\kern2pt}%
  \let\@fs@iftopcapt\iftrue}
\makeatother

\makeatletter
\def\section{\@startsection {section}{1}{\z@}{-3.25ex plus -1ex minus
 -.2ex}{1.5ex plus .2ex}{\large\bf}}
\def\subsection{\@startsection{subsection}{2}{\z@}{-3.25ex plus -1ex minus
 -.2ex}{1.5ex plus .2ex}{\normalsize\bf}}
\def\subsubsection{\@startsection{subsubsection}{3}{\z@}{3.25ex plus
 1ex minus .2ex}{-1em}{\normalsize\bf}}
\def\paragraph{\@startsection{paragraph}{4}{\z@}{3.25ex plus 1ex minus
  .2ex}{-1em}{\normalsize\bf}}
\def\subparagraph{\@startsection{subparagraph}{4}{\parindent}{3.25ex
  plus 1ex minus .2ex}{-1em}{\normalsize\bf}}
\makeatother

\makeatletter
\renewcommand{\footnotesize}{\@setfontsize\footnotesize\@viiipt{9.5}}
\makeatother

\newcommand{\smallsec}[1]{\smallskip\noindent{\bf #1.}}
\newcommand{\smallsectwo}[1]{\smallskip\noindent{\bf #1}}

\renewcommand{\paragraph}{\smallsec}

\newcommand{\seq}[1]{\langle #1 \rangle}

\newcommand{\cpachecker}{{\small\sc CPAchecker}\xspace}
\newcommand{\cpaabelf}{{\small\sc cpa-pred}\xspace}
\newcommand{\cpaabelfexp}{{\small\sc cpa-expl-pred}\xspace}
\newcommand{\cpamemo}{{\small\sc cpa-memo}\xspace}
\newcommand{\cpaexp}{{\small\sc cpa-expl}\xspace}
\newcommand{\cpaexpintpol}{{\small\sc cpa-expl}{\tiny itp}\xspace}
\newcommand{\cpaexpintpoltable}{{\small\sc cpa-expl}{\tiny itp}\xspace}
\newcommand{\cpaexpintpolabelf}{{\small\sc cpa-expl}{\tiny itp}{\small\sc -pred}\xspace}
\newcommand{\cpaexpintpolabelftable}{{\small\sc cpa-expl}{\tiny itp}{\small\sc -pred}\xspace}

\newcommand{\cseq}{{\gamma}}
\newcommand{\cseqintpol}{{\Gamma}}
\newcommand{\blasttable}{{\small\sc blast}\xspace}
\newcommand{\satabstable}{{\small\sc satabs}\xspace}

\newcommand{\blast}{{\small\sc Blast}\xspace}

\newcommand{\satabs}{{\small\sc SATabs}\xspace}

\newcommand{\spin}{{\small\sc Spin}\xspace}
\newcommand{\pathFinder}{{\small\sc Java PathFinder}\xspace}
\newcommand{\vinta}{{\small\sc Vinta}\xspace}
\newcommand{\toolDagger}{{\small\sc Dagger}\xspace}

\newcommand{\mathsat}{{\small\sc MathSAT}\xspace}
\newcommand{\smtinterpol}{{\small\sc SMTInterpol}\xspace}

\newcommand{\CPA}{{CPA}\xspace}

\newcommand{\safe}{{\sc safe}\xspace}
\newcommand{\unsafe}{{\sc unsafe}\xspace}

\newcommand{\false}{{\it false}}

\newcommand{\locs}{\mathit{L}}
\newcommand{\pc}{\mathit{l}}
\newcommand{\pci}{{\pc_0}}
\newcommand{\pct}{{\pc_{e}}}

\newcommand{\concr}{{\cal C}}
\newcommand{\sem}[1]{\ensuremath{[\![#1]\!]}\xspace}
\newcommand{\op}{\mathit{op}}
\newcommand{\SP}[2]{{\sf SP}_{#1}({#2})}
\renewcommand{\path}{\sigma}

\newcommand{\Ints}{\mathbb{Z}}
\newcommand{\varAssignment}{v}

\newcommand{\pto}{\mathrel{\longrightarrow\hspace{-4.5mm}\circ\hspace{2mm}}}
\renewcommand{\implies}{\mathrel{\Rightarrow}}
\newcommand{\defran}{\textrm{def}}

\newcommand{\interpret}[2]{{#1_{/#2}}}
\newcommand{\eval}[2]{{#1_{/#2}}}

\newcommand{\cegar}{{CEGAR}\xspace}

\newcommand{\cpaSymbol}{\mathbb{D}}

\newcommand{\transconc}[1]{\smash{\stackrel{#1}{\rightarrow}}}
\newcommand{\transabs}[2]{\smash{\stackrel{#1}{\rightsquigarrow}}}
\newcommand{\merge}{\mathsf{merge}}
\newcommand{\stopop}{\mathsf{stop}}
\newcommand{\precfn}{\mathsf{prec}}
\newcommand{\precisions}{\Pi}
\newcommand{\PREC}{\precisions}
\newcommand{\pr}{\pi}
\newcommand{\wait}{\mathsf{waitlist}}
\newcommand{\reached}{\mathsf{reached}}
\newcommand{\conccpa}{\mathbb{C}}
\newcommand{\less}{\sqsubseteq}
\newcommand{\join}{\sqcup}
\newcommand{\intlat}{\mathcal{Z}}

\begin{document}

\newcommand{\mytitle}{Explicit-Value Analysis \\ Based on CEGAR and Interpolation}

\pagestyle{empty}
\begin{minipage}{17cm}
\begin{center}
~\\[3cm]
\Huge{\mytitle}
\\[2cm]
\large{Dirk Beyer and Stefan Löwe}
\\[1cm]
\normalsize
{University of Passau, Germany}\\[7cm]

\hspace{-5mm}
\includegraphics[scale=0.2]{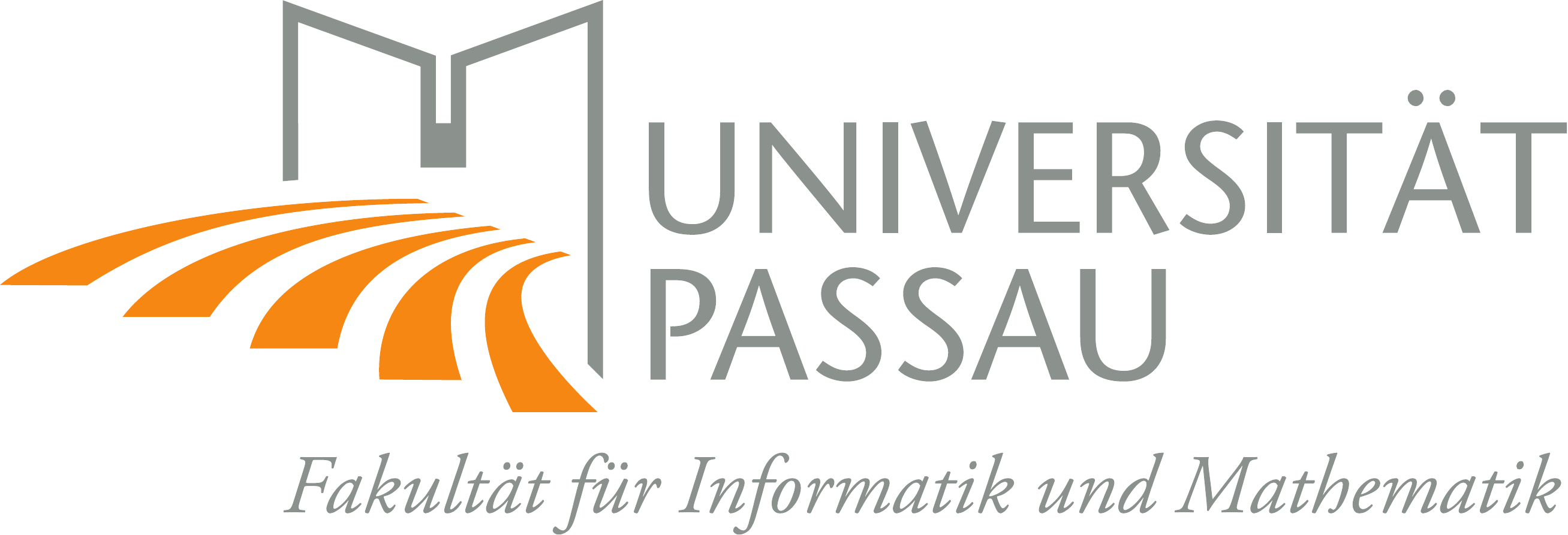} \\[1cm]
Technical Report, Number MIP-1205\\
Department of Computer Science and Mathematics\\
University of Passau, Germany\\
December 2012
\end{center}
\end{minipage}

\title{\mytitle}

\author{{Dirk Beyer and Stefan Löwe}
\vspace{3mm}\\
{University of Passau, Germany}\\
}

\maketitle
\thispagestyle{empty}
\setcounter{page}{1}
\pagestyle{plain}

\begin{abstract}
	Abstraction, counterexample-guided refinement, and interpolation
are techniques that are essential to the success of predicate-based program analysis.
These techniques have not yet been applied together to explicit-value program analysis.
We present an approach
that integrates abstraction and interpolation-based refinement into an explicit-value analysis,
i.e., a program analysis that tracks explicit values for a specified set of variables (the precision).
The algorithm uses an abstract reachability graph as central data structure
and a path-sensitive dynamic approach for precision adjustment.
We evaluate our algorithm on the benchmark set
of the Competition on Software Verification 2012 (SV-COMP'12) to
show that our new approach is highly competitive.
In addition, we show that combining our new approach with an auxiliary predicate analysis
scores significantly higher than the SV-COMP'12 winner.
\end{abstract}

\section{Introduction}
\label{section:introduction}

Abstraction is one of the most important techniques to
successfully verify industrial-scale program code,
because the abstract model omits details about the concrete semantics of the program
that are not necessary to prove or disprove the program's correctness.
Counterexample-guided abstraction refinement (\cegar)~\cite{ClarkeCEGAR} is a technique that 
iteratively refines an abstract model using counterexamples.
A counterexample is a witness of a property violation.
In software verification, the counterexamples are error paths,
i.e., paths through the program that violate the property.
\cegar starts with the most abstract model and checks if an error path can be found.
If the analysis of the abstract model does not find an error path,
then the analysis terminates, reporting that no violation exists.
If the analysis finds an error path, the path is checked for feasibility,
i.e., if the path is executable according to the concrete program semantics.
If the error path is feasible,
the analysis terminates, reporting the violation of the property,
together with the feasible error path as witness.
If the error path is infeasible,
the violation is due to a too coarse abstract model
and the infeasible error path is used to automatically refine 
the current abstraction.
Then the analysis proceeds.
Several successful tool implementations for software verification
are based on abstraction and CEGAR 
(cf.~\cite{SLAM, BLAST, MAGIC, SATABS, ARMC, CPACHECKER}).
Craig interpolation is a technique from logics that yields for two
contradicting formulas an interpolant that contains less information
than the first formula, but still enough to contradict
the second formula~\cite{Craig57}.
In software verification, interpolation can be used
to extract information from infeasible error paths~\cite{AbstractionsFromProofs},
where the resulting interpolants are used to refine the abstract model.
Predicate abstraction is a successful
abstraction technique for software model checking~\cite{GrafSaidi97},
because its symbolic state representation blends well with strongest post-conditions,
and abstractions can be computed efficiently with solvers for 
satisfiability modulo theories (SMT)~\cite{BPR01}.
\cegar and lazy refinement~\cite{LazyAbstraction}
together with interpolation~\cite{AbstractionsFromProofs}
effectively refine abstract models in the predicate domain.
The recent competition on software verification (SV-COMP'12~\cite{SVCOMP12}, Table~3)
shows that these advancements had a strong impact on the success 
of participating tools (cf.~\cite{BLAST, CPACHECKER, ARMC, CPACHECKERMEMO-COMP}).

Despite the success of abstraction, CEGAR, and interpolation 
in the field of predicate analysis,
these techniques have not yet been combined and applied together to explicit-value analysis.
We integrate these three techniques into an explicit-value analysis,
a rather unsophisticated analysis that tracks for each program variable its current value explicitly
(like constant propagation~\cite{DragonBook}, but without join).
First, we have to define the notion of abstraction for the explicit-value domain,
and the precision of the analysis (i.e., the level of abstraction)
by a set of program variables that the analysis has to track.
Second, in order to automatically determine the necessary precision
(i.e., a \emph{small} set of program variables that \emph{need} to be tracked)
we use \cegar iterations to discover finer precisions from infeasible error paths.
Third, we define interpolation for the explicit-value domain
and use this idea to construct an algorithm that efficiently extracts such a parsimonious precision
that is sufficient to eliminate infeasible error paths.

\smallsectwo{Example.}
Consider the simple example program in Fig.~\ref{fig:refineFast}.
This program contains a $while$ loop in which a system call occurs.
The loop exits if either the system call returns $0$
or a previously specified number of iterations $x$ was performed.
Because the body of the function $system\_call$ is unknown,
the value of $result$ is unknown. 
Also, the assumption $[ticks > x]$ cannot be evaluated to $true$, because $x$ is unknown.
This program is correct, i.e., the error location in line~10 is not reachable.
However, a simple explicit-value model checker that always tracks every variable
would unroll the loop, always discovering new states,
as the expression $ticks  = ticks + 1$ repeatedly assigns new values to variable~$ticks$.
Thus, due to extreme resource consumptions,
the analysis would not terminate within practical time and memory limits,
and is bound to give up on proving the safety property, eventually.

\begin{figure}[t]
\hspace{4mm}
\begin{minipage}[t]{75mm}
\begin{lstlisting}[language=C,numbersep=2pt]
extern int system_call();
int main(int x) {
  int flag, ticks, result;
  flag = 0; ticks = 0;
  while(1) {
    ticks  = ticks + 1;
    result = system_call();
    if(result == 0 || ticks > x) { break; }
  }
  if (flag > 0) { ERROR: return 1; }
}
\end{lstlisting}
\end{minipage}
\caption{Example program to illustrate the effectiveness of CEGAR-based explicit-value analysis} 
\label{fig:refineFast}
\vspace{-2mm}
\end{figure}

The new approach for explicit-value analysis that we propose can efficiently prove 
this program safe,
because it tracks only those variables that are necessary to refute the infeasible error paths.
In the first \cegar iteration,
the precision of the analysis is empty,
i.e., no variable is tracked.
Thus, the error location will be reached.
Now, using our interpolation-inspired method to discover precisions from counterexample paths,
the algorithm identifies that the variable~$flag$ (more precisely, the constraint~$flag = 0$) has
to be tracked.
The analysis is re-started after this refinement.
Because $ticks$ is not in the precision (the variable is not tracked),
the assignment $ticks = ticks + 1$ will not add new abstract states.
Since no new successors are computed,
the analysis stops unrolling the loop.
The assume operation $[flag > 0]$ 
is evaluated to $\false$, and thus, the error label is not reachable.
The analysis terminates, proving the program correct.

In summary, the crucial effect of this approach is that only relevant variables are tracked in the analysis,
while unimportant information is ignored.
This greatly reduces the number of abstract states to be visited.

\smallsec{Contributions}
We make the following contributions:
\begin{itemize}
	\item We integrate the concepts of abstraction, CEGAR, and lazy abstraction refinement 
	       into explicit-value analysis.
	\item Inspired by Craig interpolation for predicate analysis,
	       we define a novel interpolation-like approach for discovering relevant 
	       variables for the explicit-value domain.
	       This refinement algorithm is completely self-contained,
           i.e., independent from external libraries such as SMT solvers.
	\item To further improve the effectiveness and efficiency of the analysis,
	       we design a combination with a predicate analysis based 
	       on dynamic precision adjustment~\cite{CPAplus}.
	\item We provide an open-source implementation of all our concepts and give evidence of the significant improvements
	       by evaluating several approaches on benchmark verification tasks 
	       (C programs) from SV-COMP'12.
\end{itemize}

\smallsec{Related Work} 
The explicit-state model checker \spin~\cite{SPIN} 
can verify models of programs written in a language called Promela.
For the verification of C programs, tools like {\sc Modex}\,\footnote{\url{http://cm.bell-labs.com/cm/cs/what/modex/}} 
can extract Promela models from C source code.
This process requires to give a specification of the abstraction level 
(user-defined extraction rules),
i.e., the information of what should be included in the Promela model.
\spin does not provide lazy-refinement-based \cegar.
%
\pathFinder~\cite{PathFinder} is an explicit-state model checker for Java programs. 
There has been work~\cite{PathFinderWithCegar} on integrating \cegar into \pathFinder,
using an approach different from interpolation.

Program analysis with dynamic precision adjustment~\cite{CPAplus}
is an approach to adjust the precision of combined analyses on-the-fly,
i.e., during the analysis run; the precision of one analysis can be increased
based on a current situation in another analysis.
For example, if an explicit-value analysis stores too many different values
for a variable, then the dynamic precision adjustment
can remove that variable from the precision of the
explicit-value analysis and add a predicate about that variable 
to the precision of a predicate analysis.
This means that the tracking of the variable is ``moved'' from the explicit domain to the symbolic domain.
One configuration that we present later in this paper uses 
this approach (cf. \ref{smallsec:aux_pred_analysis}).

The tool \toolDagger \cite{DAGGER} improves the verification of C programs by applying
interpolation-based refinement to octagon and polyhedra domains.
To avoid imprecision due to widening in the join-based data-flow analysis,
\toolDagger replaces the standard widen operator by a so called \emph{interpolated-widen} operator,
which increases the precision of the data-flow analysis and thus avoids false alarms.
The algorithm \vinta~\cite{VINTA} applies interpolation-based refinement to
interval-like abstract domains.
If the state exploration finds an error path,
then \vinta performs a feasibility check using bounded model checking (BMC), 
and if the error path is infeasible,
it computes interpolants.
The interpolants are used to refine the invariants
that the abstract domain operates on. 
\vinta requires an SMT solver for feasibility checks and interpolation.

More tools are mentioned in our evaluation section, where we
compare (in terms of precision and efficiency) our tool implementation with tools
that participated in SV-COMP'12.

There is, to the best of our knowledge, no work that
integrates abstraction, \cegar, lazy refinement, and interpolation into 
explicit-state model checking.
We make those techniques available for the explicit-value domain.

\section{Preliminaries}
\label{section:preliminaries}

Our approach is based on several existing concepts,
and in this section we remind the reader of 
some basic definitions.

\subsection{Programs, Control-Flow Automata, States}
We restrict the presentation to
a simple imperative programming language,
where all operations are either assignments or assume operations,
and all variables range over integers\,%
\footnote{Our implementation is based on \cpachecker{},
which operates on C~programs;
non-recursive function calls are supported.}.
The following definitions are taken from previous work~\cite{ABE}:
A program is represented by a \emph{control-flow automaton} CFA.
A CFA $A = (\locs, G)$ consists of
a set~$\locs$ of program locations, which model the program counter,
and a set $G \subseteq \locs \times Ops \times \locs$ of control-flow edges,
which model the operations that are executed when
control flows from one program location to another.
The set of program variables that occur in operations from~$Ops$
is denoted by~$X$.
A~\emph{verification problem}~${P = (A, \pci, \pct)}$ consists of 
a CFA $A$, representing the program,
an initial program location~$\pci \in \locs$, representing the program entry, and
a target program location~$\pct \in \locs$, which represents the error.

A \emph{concrete data state} of a program is
a variable assignment $cd: X \to \Ints$,
which assigns to each program variable an integer value.
A~\emph{concrete state} of a program is a pair~$(l, cd)$,
where $l \in \locs$ is a program location and $cd$~is a concrete data state.
The set of all concrete states of a program is denoted by~$\concr$,
a subset~$r \subseteq \concr$ is called \emph{region}.
Each edge~$g \in G$ defines a labeled transition relation 
$\mathord{\transconc{g}} \subseteq \concr \times \{g\} \times \concr$.
The complete transition relation~$\transconc{}$ is the union over 
all control-flow edges:
$\mathord{\transconc{}} = \bigcup_{g \in G} \transconc{g}$.
We write $c \transconc{g} c'$ if $(c, g, c') \in \mathord{\transconc{}}$,
and $c \transconc{} c'$ if there exists a $g$ with $c \transconc{g} c'$.

An \emph{abstract data state} represents a region of concrete data states,
formally defined as abstract variable assignment.
An \emph{abstract variable assignment} is 
a partial function~${v: X \pto \Ints \cup \{\top, \bot\}}$,
which maps variables in the definition range of function~$v$ to
integer values or~$\top$ or~$\bot$.
The special value~$\top$ is used to represent an unknown value, e.g.,
resulting from an uninitialized variable or an external function call,
and the special value~$\bot$ is used to represent no value, 
i.e., a contradicting variable assignment.  
We denote the \emph{definition range} for a partial function~$f$ as 
${\defran(f) = \{ x \mid \exists y: (x, y) \in f\}}$,
and the \emph{restriction} of a partial function~$f$ 
to a new definition range~$Y$ as~$f_{|Y} = {f \cap (Y \times (\Ints \cup \{\top, \bot\}))}$.
An abstract variable assignment~$v$ represents 
the region~$\sem{v}$ of all concrete data states~$cd$ 
for which $\varAssignment$ is valid,
formally: $\sem{v} = {\{cd \mid \forall x \in \defran(v): 
cd(x) = v(x)}$ or $v(x) = \top\}$.
An~\emph{abstract state} of a program is a pair~$(l, \varAssignment)$,
representing the following set of concrete states:
$\{ (l, cd) \mid cd \in \sem{v}\}$.

\subsection{Configurable Program Analysis with \\ Dynamic Precision Adjustment}
\label{sec:cpap-def}

We use the framework of configurable program analysis (\CPA)~\cite{CPA},
extended by the concept of dynamic precision adjustment~\cite{CPAplus}.
Such a \CPA supports adjusting the precision of an analysis
during the exploration of the program's abstract state space.
A \textit{composite} \CPA{} can control the precision of its 
component analyses during the verification process, i.e.,
it can make a component analysis more abstract, and thus more efficient,
or it can make a component analysis more precise, and thus more expensive.
A \CPA{}
$\cpaSymbol = (D, \precisions, \transabs{}{}, \merge, \stopop, \precfn)$
consists of 
(1)~an abstract domain~$D$,
(2)~a set~$\precisions$ of precisions,
(3)~a transfer relation~$\transabs{}{}{}$,
(4)~a merge operator~$\merge$,
(5)~a termination check~$\stopop$, and
(6)~a precision adjustment function~$\precfn$.
Based on these components and operators, we can formulate a flexible and customizable
reachability algorithm, which is adapted from previous work~\cite{CPA,FMCAD12}.

\subsection{Explicit-Value Analysis as CPA\label{sec:cPCPA}}

In the following, we define a component CPA
that tracks explicit integer values for program variables.
In order to obtain a complete analysis,
we construct a composite \CPA that consists of the component CPA 
for explicit values and another component CPA 
for tracking the program locations
(\CPA for location analysis, as previously described~\cite{CPAplus}).
For the composite \CPA,
the general definitions of the abstract domain, the transfer relation, 
and the other operators
are given in previous work~\cite{CPAplus};
the composition is done automatically by the framework implementation \cpachecker.

The \emph{\CPA for explicit-value analysis},
which tracks integer values for the variables of a program explicitly,
is defined as 
$\conccpa = (D_\conccpa, \precisions_\conccpa, \transabs{}{}_\conccpa, 
 \merge_\conccpa, \stopop_\conccpa, \precfn_\conccpa)$
and consists of the following components~\cite{CPAplus}:

\smallsec{1}
The abstract domain~$D_\conccpa = (C, \mathcal{V}, \sem{\cdot})$ 
contains the set $C$ of concrete data states,
and uses the semi-lattice~$\mathcal{V} = {(V, \top, \bot, \less, \join)}$, 
which consists of the set~$V = (X \pto \intlat)$ of abstract variable assignments,
where $\intlat = \Ints \cup \{\top_\intlat, \bot_\intlat\}$ 
induces the flat lattice over the integer values
(we write~$\Ints$ to denote the set of integer values).
The top element $\top \in V$, with ${\top(x) = \top_\intlat}$ for all $x \in X$,
is the abstract variable assignment that holds no specific value for any variable,
and the bottom element $\bot \in V$, with $\bot(x) = \bot_\intlat$ for all $x \in X$,
is the abstract variable assignment which models that there is no value assignment possible,
i.e., a state that cannot be reached in an execution of the program.
The partial order~$\mathord{\less} \subseteq V \times V$ is defined as
${v \less v'}$ if for all $x \in X$, we have
   $v(x) = v'(x)$ or $v(x) = \bot_\intlat$ or $v'(x) = \top_\intlat$. 
The join $\join: V \times V \to V$ yields the least upper bound
for two variable assignments.
The concretization function ${\sem{\cdot}: V \to 2^C}$
assigns to each abstract data state~$v$ its meaning, i.e., 
the set of concrete data states
that it represents.

\smallsec{2}
The set of precisions~$\precisions_\conccpa = 2^X$ 
is the set of subsets of program variables.
A precision~$\pr \in \precisions_\conccpa$ 
specifies a set of variables to be tracked.
For example, $\pr = \emptyset$ means that no variable
is tracked, and 
$\pr = X$ means that every program variable
is tracked.

\smallsec{3}
The transfer relation~$\transabs{}{}_\conccpa$ has the transfer
$v \transabs{g}{} (v',\pr)$ if\\
(1) $g = (\cdot, \mathtt{assume}(p), \cdot)$ and
for all $x \in X:$\\ 
$v'(x) = \left\{
\begin{array}{ll}
  \bot_\intlat & \text{if } (y, \bot_\intlat) \in v \text{ for some } y \in X  \\
                 & \text{ or the formula } \interpret{p}{v} 
                   \text{ is unsatisfiable}\\
  c              & \text{if } c \text{ is the only satisfying assignment of} \\
                 &  \text{ the formula } \interpret{p}{v} \text{ for variable } x\\ 
  \top_\intlat & \text{otherwise}
\end{array}
\right.$\\
where $\interpret{p}{v}$ denotes the interpretation of a predicate~$p$ over variables from~$X$
for an abstract variable assignment~$v$,
that is, 
$\interpret{p}{v} =\\
p 
~\land~ 
\bigwedge\limits_{x \in \defran(v), v(x) \in \Ints} x = v(x)
~\land~ 
\lnot\exists x \in \defran(v): v(x) = \bot_\intlat$
\\[0.5ex]
or\\[0.5ex]
(2) $g = (\cdot, \mathtt{ w:=\, } exp, \cdot)$ and
for all $x \in X:$\\
$v'(x) = \left\{
\begin{array}{ll}
  \eval{exp}{v}       & \text{if } x = \mathtt{w} \\
  v(x)                 & \text{if } x \in \defran(v) \\
  \top_\intlat & \text{otherwise}
\end{array}
\right.$\\
where $\eval{exp}{v}$ denotes the interpretation of an expression~$exp$ over variables from~$X$ 
for an abstract value assignment~$v$:\\
$\eval{exp}{v} =
\left\{ \begin{array}{ll}
  \bot_\intlat & \text{if } (y, \bot_\intlat) \in v \text{ for some } y \in X \\
  \top_\intlat & \text{if } (y, \top_\intlat) \in v \text{ or } y \not\in \defran(v) \\
          & \text{ for some } y \in X \text{ that occurs in } exp\\
  c       & \text{otherwise, where expression $exp$} \\
          & \text{ evaluates to $c$ after replacing each} \\
          & \text{ occurrence of variable~$x$ with $x \in \defran(v)$ } \\
          & \text{ by~$v(x)$ in $exp$ }
\end{array} \right.$

\smallsec{4}
  The merge operator does not combine elements when control flow meets:
  $\merge_\conccpa(v, v', \pr) = v'$.

\smallsec{5}
  The termination check considers abstract states individually:
  $\stopop_\conccpa(v, R, \pr) = (\exists v' \in R: v \less v')$.

\smallsec{6}
The precision adjustment function computes a new abstract state with precision
based on the abstract state $v$ and the precision $\pr$ by restricting the variable assignment~$v$
to those variables that appear in $\pr$, formally:
$\precfn(v,\pr,R) = (v_{|\pr},\pr)$.
(In this analysis instance, $\precfn$ only adjusts the abstract state
according to the current precision~$\pr$,
and leaves the precision itself unchanged.)

The precision of the analysis controls which program variables are tracked 
in an abstract state.
In other approaches, this information is hard-wired in either 
the abstract-domain elements or the algorithm itself.
The concept of \CPA supports different precisions for different abstract states.
A simple analysis can start with an initial precision
and propagate it to new abstract states, such that the overall analysis
uses a globally uniform precision.
It is also possible to specify a precision individually per program location,
instead of using one global precision.
Our refinement approach in the next section will be based on location-specific precisions.

\subsection{Predicate Analysis as CPA}

The abstract domain of predicates~\cite{GrafSaidi97} was successfully
used in several tools for software model checking 
(e.g.,~\cite{SLAM, BLAST, MAGIC, SATABS, ARMC, CPACHECKER}).
In a predicate analysis, the precision is defined as a set of predicates,
and the abstract states track the strongest set of predicates that
are fulfilled (cartesian predicate abstraction) or the strongest
boolean combination of predicates that are fulfilled (boolean predicate abstraction).
This means, the abstraction level of the abstract model is determined
by predicates that are tracked in the analysis.
Predicate analysis is also implemented as a CPA in the framework~\cpachecker,
and a detailed description is available~\cite{ABE}.
The precision is freely adjustable also in the predicate analysis,
and we use this feature later in this article to compose a combined analysis.
This analysis uses the predicate analysis to track variables that have many distinct
values --- a scenario in which the explicit-value analysis alone would be inefficient.
The combined analysis adjusts the overall precision by removing
variables with many distinct values from the precision of the explicit-value analysis
and adds predicates about these variables to the precision of the predicate analysis~\cite{CPAplus}
to allow the combined analysis to run efficiently.

\subsection{Lazy Abstraction}

The concept of lazy abstraction~\cite{LazyAbstraction} consists of two ideas:
First, the abstract reachability graph (ARG) ---the unfolding of the control-flow graph,
representing our central data structure to store abstract states---
is constructed on-the-fly, 
i.e., only when needed and only for parts of the state space that are reachable.
We implement this using the standard reachability algorithm for CPAs as described in the
next subsection.
Second, the abstract states in the ARG are refined only where necessary
along infeasible error paths in order to eliminate those paths.
This is implemented by using CPAs with dynamic precision adjustment,
where the refinement procedure operates on location-specific precisions
and where the precision-adjustment operator always removes unnecessary
information from abstract states, as outlined above.

\subsection{Reachability Algorithm for CPA}

Algorithm~\ref{algo:analysis} keeps updating two sets of abstract states with precision:
the set $\reached$ to store all abstract states with precision that are found to be 
reachable, and a set $\wait$ to store all abstract states with precision that are not 
yet processed, i.e., the frontier.
The state exploration starts with choosing and removing an abstract state with precision from the $\wait$,
and the algorithm considers each abstract successor according to the transfer relation.
Next, for the successor, the algorithm adjusts the precision of the successor
using the precision adjustment function~$\precfn$.
If the successor is a target state
(i.e., a violation of the property is found),
then the algorithm terminates, 
returning the current sets $\reached$ and $\wait$
--- possibly as input for a subsequent precision refinement, 
as shown below (cf. Alg.~\ref{algorithm:cegar}).
Otherwise, using the given operator~$\merge$,
the abstract successor state is combined with each existing abstract state from~$\reached$.
If the operator~$\merge$ has added information to the new abstract state,
such that the old abstract state is subsumed, then 
the old abstract state with precision is replaced by the new abstract state with precision
in the sets~$\reached$ and~$\wait$.
If after the merge step the resulting new abstract state with precision is covered by the
set~$\reached$, then further exploration of this abstract state is stopped.
Otherwise, the abstract state with its precision is added to the set~$\reached$ and to the set~$\wait$.
Finally, once the set $\wait$ is empty, the set~$\reached$ is returned.

\begin{algorithm}[t]
\begin{small}
\caption{\textbf{ }$\textsf{CPA}(\cpaSymbol, R_0, W_0)$, adapted from \cite{CPAplus}
         \label{algo:analysis}}
\begin{algorithmic}
\INPUT        a CPA $\cpaSymbol = (D, \precisions, \transabs{}{}, \merge, \stopop, \precfn)$,\\
\hspace{6mm}  a set~$R_0 \subseteq (E \times \precisions)$ of abstract states with precision,\\
\hspace{6mm}  a subset~$W_0 \subseteq R_0$ of frontier abstract states with precision,\\
\hspace{6mm}  where $E$ denotes the set of elements of the semi-lattice of $D$
\OUTPUT       a set of reachable abstract states with precision, \\
\hspace{6mm}  a subset of frontier abstract states with precision
\VARDECL      two sets $\reached$ and $\wait$ of elements of $E \times \precisions$

\STATE $\reached := R_0$;
$\wait := W_0$;
\WHILE {$\wait \not= \emptyset$}
  \STATE choose $(e, \pr)$ from $\wait$; remove $(e, \pr)$ from $\wait$;
  \FOR{each $e'$ with $e \transabs{}{} (e',\pr)$}
  	\STATE // Precision adjustment.
	\STATE $(\hat{e},\hat{\pr}) := \precfn(e',\pr,\reached)$;
    \IF {$\mathsf{isTargetState}(\hat{e})$}
      \STATE {\bf return } $\big(\reached \cup (\hat{e}, \hat{\pr}), \wait\big)$;
    \ENDIF
    \FOR{each $(e'', \pr'') \in \reached$}
      \STATE // Combine with existing abstract state.
      \STATE $e_{new} := \merge(\hat{e}, e'', \hat{\pr})$;
      \IF {$e_{new} \not= e''$}
        \STATE $\wait    := \big(\wait    \cup \{(e_{new},\hat{\pr})\}\big) \setminus \{(e'',\pr'')\}$;
        \STATE $\reached := \big(\reached \cup \{(e_{new},\hat{\pr})\}\big) \setminus \{(e'',\pr'')\}$;
      \ENDIF
    \ENDFOR
    \STATE // Add new abstract state?
    \IF {$\lnot~\mathord{\stopop}\big(\hat{e}, \big\{ e \mid (e,\cdot) \in \reached\big\}, \hat{\pr}\big)$}
      \STATE $\wait := \wait \cup \{(\hat{e},\hat{\pr})\}$;  \\
             $\reached := \reached \cup \{(\hat{e},\hat{\pr})\}$
    \ENDIF
  \ENDFOR
\ENDWHILE
\STATE {\bf return } $(\reached, \emptyset)$;
\end{algorithmic}
\end{small}
\end{algorithm}


\subsection{Counterexample-Guided Abstraction Refinement}
\label{sec:cegar}

Counterexample-guided abstraction refinement (CEGAR)~\cite{ClarkeCEGAR}
is a technique for automatic stepwise refinement of an abstract model.
CEGAR is based on three concepts:
(1) a \emph{precision}, which determines the current level of abstraction,
(2) a \emph{feasibility check}, deciding if an abstract error path is feasible,
    i.e., if there exists a corresponding concrete error path, and
(3) a \emph{refinement} procedure, which takes as input an infeasible
    error path and extracts a precision that suffices to
    instruct the exploration algorithm to not explore the same path again later.
Algorithm \ref{algorithm:cegar} shows an outline of a generic and simple CEGAR algorithm.
The algorithm starts checking a program using a coarse initial \emph{precision}~$\pr_0$.
It uses the reachability algorithm Alg.~\ref{algo:analysis}
for computing the reachable abstract state space, returning the sets $\reached$ and $\wait$.
If the analysis has exhaustively checked all program states and did not reach the error,
indicated by an empty set~$\wait$,
then the algorithm terminates and reports that the program is safe.  
If the algorithm finds an error in the abstract state space,
i.e., a counterexample for the given specification,
then the exploration algorithm stops and returns the unfinished, incomplete
sets~$\reached$ and $\wait$.
Now the according abstract error path is extracted 
from the set~$\reached$ using procedure~$\mathsf{extractErrorPath}$
and analyzed for feasibility
using the procedure~$\mathsf{isFeasible}$ for \emph{feasibility check}.
If the abstract error path is feasible,
meaning there exists a corresponding concrete error path, 
then this error path represents a violation of the specification and 
the algorithm terminates, reporting a bug. 
If the error path is infeasible, i.e., not corresponding to a concrete program path,
then the precision was too coarse and needs to be refined.
The algorithm extracts certain information from the error path
in order to refine the precision based on that information using 
the procedure~$\mathsf{Refine}$ for \emph{refinement},
which returns a precision~$\pr$ that makes the analysis strong enough to refute the
infeasible error path in further state-space explorations.
The current precision is extended using the precision returned by the refinement procedure
and the analysis is restarted with this refined precision.
Instead of restarting from the initial sets for $\reached$ and $\wait$,
we can also prune those parts of the ARG that need to be rediscovered
with new precisions, and replace the precision of the leaf nodes in the ARG
with the refined precision, and then restart the exploration on the pruned sets.
Our contribution in the next section is
to introduce new implementations for the feasibility check as well as for the refinement procedure.

\begin{algorithm}[t]
\begin{small}
\caption{\textbf{ }$\textsf{CEGAR}(\cpaSymbol, e_0, \pr_0)$
         \label{algorithm:cegar}}
\begin{algorithmic}
\INPUT              a configurable program analysis with dynamic precision\\
\hspace{6mm}       adjustment
                     $\cpaSymbol = (D, \precisions, \transabs{}{}, \merge, \stopop, \precfn)$,\\
\hspace{6mm}       an initial abstract state~$e_0 \in E$ with precision~$\pr_0 \in \precisions$,\\
\hspace{6mm}       where $E$ denotes the set of elements of the semi-lattice of $D$
\OUTPUT \hspace{1.5mm} verification result $\mathit{\safe}$ or $\mathit{\unsafe}$
\STATE {\hspace{-3mm}\bf Variables:} a set $\reached$ of elements of $E \times \precisions$,\\
\hspace{11mm}                       a set $\wait$ of elements of $E \times \precisions$,\\
\hspace{11mm}                       an error path $\path = \seq{(\op_1, \pc_1), ..., (\op_{n}, \pc_{n})}$

\STATE $\reached := \{(e_0, \pr_0)\}$;
$\wait := \{(e_0, \pr_0)\}$;
$\pr := \pr_0$;
\WHILE {$true$}
  \STATE $(\reached, \wait) := \textsf{\CPA}(\cpaSymbol, \reached, \wait)$;
  \IF {$\wait = \emptyset$}
    \STATE {\bf return } $\mathit{\safe}$
  \ELSE
    \STATE $\path := \mathsf{extractErrorPath}(\reached)$;
    \IF [error path is feasible: report bug] {$\mathsf{isFeasible}(\path)$}
      \STATE {\bf return } $\mathit{\unsafe}$
    \ELSE [error path is not feasible: refine and restart] 
      \STATE $\pr := \pr \cup \mathsf{Refine}(\path)$;
      \STATE $\reached := {(e_0, \pr)}$;
             $\wait := {(e_0, \pr)}$;
    \ENDIF
  \ENDIF 
\ENDWHILE
\end{algorithmic}
\end{small}
\end{algorithm}

\subsection{Interpolation}

For a pair of formulas $\varphi^-$ and $\varphi^+$
such that $\varphi^- \land \varphi^+$ is unsatisfiable,
a Craig interpolant $\psi$ is a formula that fulfills the following requirements~\cite{Craig57}:
\begin{enumerate}
	\item the implication $\varphi^- \Rightarrow \psi$ holds,
	\item the conjunction $\psi \land \varphi^+$ is unsatisfiable, and
	\item $\psi$ only contains symbols that occur in both $\varphi^-$ and $\varphi^+$.
\end{enumerate}
Such a Craig interpolant is guaranteed to exist 
for many useful theories, for example,
the theory of linear arithmetic with uninterpreted functions,
as implemented in some SMT solvers 
(e.g., \mathsat%
\footnote{\url{http://mathsat4.disi.unitn.it}}, 
\smtinterpol%
\footnote{\url{http://ultimate.informatik.uni-freiburg.de/smtinterpol}}).

\cegar based on Craig interpolation has been proven successful in the predicate domain.
Therefore, we investigate if this technique is also beneficial for explicit-value model checking.
Interpolants from the predicate domain, which consist of path formulas,
are not useful for the explicit domain.
Hence, we need to develop a procedure to compute interpolants for the explicit domain,
which we introduce in the following section.

\section{Refinement-Based Explicit-Value Analysis}
\label{section:explicit_refinement}

The level of abstraction in our explicit-value analysis is determined by the precisions
for abstract variable assignments over program variables.
The CEGAR-based iterative refinement needs an extraction method to
obtain the necessary precision from infeasible error paths.
We use our novel notion of interpolation for the explicit domain
to achieve this goal.

\subsection{Explicit-Value Abstraction}

We now introduce some necessary operations on abstract variable assignments,
the semantics of operations and paths, and the precision for abstract
variable assignments and programs, in order to be able to concisely
discuss interpolation for abstract variable assignments and constraint sequences.

The operations \emph{implication} and \emph{conjunction} for 
abstract variable assignments are defined as follows:
implication for $v$ and~$v'$:
$v \implies v'$ if $\defran(v') \subseteq \defran(v)$ and 
for each variable $x \in \defran(v) \cap \defran(v')$
we have $v(x) = v'(x)$ or $v(x) = \bot$ or $v'(x) = \top$;
conjunction for $v$ and~$v'$: for each variable $x \in \defran(v) \cup \defran(v')$ we have\\
$(v \land v')(x) = 
\left\{
\begin{array}{ll}
    v(x)				& \textrm{if } x \in \defran(v) \textrm{ and } x \not\in \defran(v')\\
    v'(x)				& \textrm{if } x \not\in \defran(v) \textrm{ and } x \in \defran(v')\\
    v(x)                & \textrm{if } v(x) = v'(x)\\
    \bot               	& \textrm{if } \top \not= v(x) \not= v'(x) \not= \top\\
    \top				& \textrm{otherwise } (v(x) = \top \textrm{ or } v'(x) = \top)
\end{array}
\right.$\\

Furthermore we define \emph{contradiction} for an abstract variable assignment $v$:
$v$ is contradicting if there is a variable~$x \in \defran(v)$
such that $v(x) = \bot$ (which implies $\sem{v} = \emptyset$);
and \emph{renaming} for $v$: the abstract variable assignment~$v^{x \mapsto y}$, with $y \not \in \defran(v)$,
results from $v$ by renaming variable~$x$ to~$y$:
$v^{x \mapsto y} = (v \setminus \{ (x, v(x)) \}) \cup \{(y, v(x))\}$.

The \emph{semantics of an operation} $\op \in Ops$ is defined by the
strongest post-operator~$\SP{\op}{\cdot}$ for abstract variable assignments: 
given an abstract variable assignment~$\varAssignment$, $\SP{\op}{\varAssignment}$
represents the set of data states that are reachable from 
any of the states in the region represented by $\varAssignment$ after the execution of $\op$.
Formally, given an abstract variable assignment~$v$ and an assignment operation~$s := exp$,
we have
$\SP{s := exp}{\varAssignment} = v_{|{X \setminus \{s\}}} \land v_{s := exp}$ with 
$v_{s := exp} = \{ (s, \eval{exp}{v})\}$,
where $\eval{exp}{v}$ denotes the interpretation of expression~$exp$ 
for the abstract variable assignment~$v$
(cf. definition of $\eval{exp}{v}$ in Subsection~\ref{sec:cPCPA}).
That is, the value of variable~$s$ is the result of the arithmetic evaluation of expression~$exp$,
or $\top$ if not all values in the expression are known,
or $\bot$ if no value is possible 
(an abstract data state in which a variable is assigned to $\bot$ 
 does not represent any concrete data state).
Given an abstract variable assignment~$v$
and an assume operation\,%
${[p]}$,
we have $\SP{[p]}{v} = v'$ and for all~$x \in X$ we have
$v'(x) = \bot$ if $(y, \bot) \in v$ for some variable~$x \in X$ or
the formula~$\interpret{p}{v}$ is unsatisfiable,
or 
$v'(x) = c$ if c is the only satisfying assignment of the formula~$\interpret{p}{v}$ for variable~$x$,
or
$v'(x) = \top$ in all other cases;
the formula $\interpret{p}{v}$ is defined as in Subsection~\ref{sec:cPCPA}.

A \emph{path}~$\path$ is a sequence~$\seq{(\op_1, \pc_1), ..., (\op_n, \pc_n)}$
of pairs of an operation and a location.
The path~$\path$ is called \emph{program path} if for every $i$ with $1 \leq i \leq n$
there exists a CFA edge~$g = (\pc_{i-1}, \op_i, \pc_i)$ and $\pc_0$ is the initial program location,
i.e., $\path$ represents a syntactic walk through the CFA.
Every path~$\path = \seq{(\op_1, \pc_1), ..., (\op_n, \pc_n)}$ defines a \emph{constraint sequence}
$\cseq_\path = \seq{\op_1, ..., \op_n}$.
The \emph{semantics of a program path}
$\path = \seq{(\op_1, \pc_1), ..., (\op_n, \pc_n)}$
is defined as the successive application of the strongest post-operator 
to each operation of the corresponding constraint sequence~$\cseq_\path$:
$\SP{\cseq_\path}{\varAssignment} = \SP{\op_n}{ ... \SP{\op_i}{.. \SP{\op_1}{\varAssignment} ..}... }$.
The set of concrete program states that result from running $\path$
is represented by the pair~$(l_n, \SP{\cseq_\path}{v_0})$, 
where $v_0 = \{\}$ is the initial abstract variable assignment 
that does not map any variable to a value.
A program path $\path$ is \emph{feasible} if $\SP{\cseq_\path}{v_0}$ 
is not contradicting, i.e., $\SP{\cseq_\path}{v_0}(x) \not= \bot$ 
for all variables $x$ in~$\defran(\SP{\cseq_\path}{v_0})$.
A concrete state~$(l_n, cd_n)$ is \emph{reachable} from a region~$r$,
denoted by $(l_n, cd_n) \in Reach(r)$,
if there exists a feasible program path $\path = \seq{(\op_1, \pc_1), ..., (\op_n, \pc_n)}$
with $(l_0, v_0) \in r$ and $cd_n \in \sem{\SP{\cseq_\path}{v_0}}$.
A location $l$ is reachable if there exists a concrete state $c$ such that $(l, c)$ is reachable.
A program is \safe if $\pct$ is not reachable.

The \emph{precision for an abstract variable assignment}
is a set~$\pi$ of variables.
The \emph{explicit-value abstraction} for an abstract variable assignment
is an abstract variable assignment that is defined only on variables
that are in the precision~$\pi$.
For example, the explicit-value abstraction
for the variable assignment~${v = \{x \mapsto 2, y \mapsto 5\}}$ and the
precision $\pi = \{x\}$ is the 
abstract variable assignment~$v^\pi = \{x \mapsto 2\}$.

The \emph{precision for a program} is a function~$\PREC: L \to 2^X$,
which assigns to each program location a precision for an abstract variable assignment,
i.e., a set of variables for which the analysis is instructed to track values.
A \emph{lazy explicit-value abstraction} of a program
uses different precisions for different abstract states on different program paths 
in the abstract reachability graph (ARG).
The explicit-value abstraction for a variable assignment at location~$l$
is computed using the precision~$\PREC(l)$.

\subsection{CEGAR for Explicit-Value Model Checking}
We now instantiate the three components of the CEGAR technique,
i.e., precision, feasibility check, and refinement,
for our explicit-value analysis.
The precisions that our CEGAR instance uses are the above introduced 
precisions for a program (which assign to each program location a set of variables),
and we start the CEGAR iteration with the empty precision, i.e., 
$\PREC_{init}(l) = \emptyset$ for each $l \in L$, such that no variable will be tracked.

The feasibility check for a path~$\path$ is performed by executing an explicit-value analysis of
the path~$\path$ using the
full precision~$\PREC(l) = X$ for all locations~$l$, i.e., all variables will be tracked.
This is equivalent to computing $\SP{\cseq_\path}{v_0}$ and check if the result is contradicting,
i.e., if there is a variable for which the resulting abstract variable assignment is~$\bot$.
This feasibility check is extremely efficient, because the path is finite 
and the strongest post-operations for abstract variable assignments
are simple arithmetic evaluations.
If the feasibility check reaches the error location~$\pct$,
then this error can be reported.
If the check cannot reach the error location,
because of a contradicting abstract variable assignment,
then a refinement is necessary because at least one constraint
depends on a variable that was not yet tracked.

We define the last component of the CEGAR technique, the refinement,
after we introduced the notion of interpolation for variable assignments and
constraint sequences.

\subsection{Interpolation for Variable Assignments}

For each infeasible error path in the above mentioned refinement operation,
we need to determine a precision that assigns to each program location on that path
the set of program variables that the explicit-value analysis needs to track
in order to eliminate that infeasible error path in future explorations.
Therefore, we define an interpolant
for abstract variable assignments.

\newcommand{\intpol}{{\mathcal{V}}}
An \emph{interpolant} for a pair of abstract variable assignments
$v^-$ and $v^+$, such that $v^- \land v^+$ is contradicting,
is an abstract variable assignment~$\intpol$
that fulfills the following requirements:
\begin{enumerate}
	\item the implication $v^- \implies \intpol$ holds,
	\item the conjunction $\intpol \land v^+$ is contradicting, and
	\item $\intpol$ only contains variables in its definition range
	       which are in the definition ranges of both $v^-$ and $v^+$
	       ($\defran(\intpol) \subseteq \defran(v^-) \cap \defran(v^+)$).
\end{enumerate}

\smallsec{Lemma}
For a given pair $(v^-$, $v^+)$ of abstract variable assignments, 
such that $v^- \land v^+$ is contradicting,
an interpolant exists.
Such an interpolant can be computed in time $O(m+n)$,
where $m$ and $n$ are the sizes of $v^-$ and $v^+$, respectively.

\smallsec{Proof}
The variable assignment~$v^-_{|\defran(v^+)}$
is an interpolant for the pair~$(v^-$, $v^+)$.

\smallsec{Note}
The above-mentioned interpolant that simply results
from restricting~$v^-$ to the definition range of~$v^+$ (common definition range) 
is of course not a `good' interpolant.
In practice, we strive for interpolants with minimal definition range, 
and use slightly more expensive algorithms to compute them.
Interpolation for abstract variable assignments is a first idea to approach the problem,
but since we need to extract interpolants for paths, 
we next define interpolation for constraint sequences.

\subsection{Interpolation for Constraint Sequences}
A more expressive interpolation can be achieved by considering constraint sequences.
The \emph{conjunction}~${\cseq \land \cseq'}$ of two constraint sequences
${\cseq = \seq{\op_1, ..., \op_n}}$ and $\cseq' = \seq{\op_1', ..., \op_m'}$
is defined as their concatenation, i.e., ${\cseq \land \cseq' =  \seq{\op_1, ..., \op_n, \op_1', ..., \op_m'}}$,
the \emph{implication} of $\cseq$~and~$\cseq'$ (denoted by $\cseq \implies \cseq'$) 
as $\SP{\cseq}{v_0} \implies \SP{\cseq'}{v_0}$,
and $\cseq$~is \emph{contradicting} if $\sem{\SP{\cseq}{v_0}} = \emptyset$, with $v_0 = \{\}$.

An \emph{interpolant} for a pair of constraint sequences
$\cseq^-$ and~$\cseq^+$, such that $\cseq^- \land \cseq^+$ is contradicting,
is a constraint sequence ~$\cseqintpol$
that fulfills the following requirements:
\begin{enumerate}
	\item the implication $\cseq^- \implies \cseqintpol$ holds,
	\item the conjunction $\cseqintpol \land \cseq^+$ is contradicting, and
	\item $\cseqintpol$ contains in its constraints only variables
	       that occur in the constraints of both $\cseq^-$ and $\cseq^+$.
\end{enumerate}

\smallsec{Lemma}
For a given pair $(\cseq^-$, $\cseq^+)$ of constraint sequences, 
such that $\cseq^- \land \cseq^+$ is contradicting,
an interpolant exists.
Such an interpolant is computable in time $O(m \cdot n)$,
where $m$ and $n$ are the sizes of $\cseq^-$ and $\cseq^+$, respectively.

\begin{algorithm}[t]
\begin{small}
\caption{$\mathsf{Interpolate}(\cseq^-, \cseq^+)$}
\label{algorithm:interpolate}
\begin{algorithmic} 
\INPUT{two constraint sequences $\cseq^-$ and $\cseq^+$,\\ 
        \hspace{10mm} with $\cseq^- \land \cseq^+$ is contradicting}
\OUTPUT{a constraint sequence $\cseqintpol$, \\
        \hspace{10mm} which is an interpolant for $\cseq^-$ and $\cseq^+$}
\VARDECL{an abstract variable assignment $v$}

\STATE{$v := \SP{\cseq^-}{\emptyset}$}
\FOR{{\bf each} $x \in \defran{(v)}$}
  \IF{$\SP{\cseq^+}{v_{|\defran(v) \setminus \{x\}}}$ is contradicting}
    \STATE{// $x$ is not relevant and should not occur in the interpolant}
    \STATE{$v := v_{|\defran(v) \setminus \{x\}}$}
  \ENDIF
\ENDFOR
\STATE{// construct the interpolating constraint sequence}
\STATE{$\cseqintpol := \seq{}$}
\FOR{{\bf each} $x \in \defran{(v)}$}
  \STATE{// construct an assume constraint for $x$}
  \STATE{$\cseqintpol := \cseqintpol \land \langle [x = v(x)] \rangle$}
\ENDFOR
\STATE{\bf return $\cseqintpol$}
\end{algorithmic}
\end{small}
\end{algorithm}

\smallsec{Proof}
Algorithm~$\mathsf{Interpolate}$ (Alg.~\ref{algorithm:interpolate}) returns 
an interpolant for two constraint sequences $\cseq^-$ and~$\cseq^+$.
The algorithm starts with computing the strongest post-condition for $\cseq^-$ and assigns the result
to the abstract variable assignment~$v$, which then may contain up to $m$ variables.
Per definition, the strongest post-condition for~$\cseq^+$ 
of variable assignment~$v$ is contradicting.
Next we try to eliminate each variable from~$v$,
by testing if removing it from $v$
makes the strongest post-condition for~$\cseq^+$ of~$v$ contradicting
(each such test takes $n$~$\mathsf{SP}$~steps).
If it is contradicting, the variable can be removed.
If not, the variable is necessary to prove the contradiction of the two constraint sequences,
and thus, should occur in the interpolant.
Note that this keeps only variables in~$v$ that occur in~$\cseq^+$ as well.
The rest of the algorithm constructs a constraint sequence from the variable assignment,
in order to return an interpolating constraint sequence, 
which fulfills the three requirements of an interpolant.
A naive implementation can compute such an interpolant in $O((m + n)^3)$.

\subsection{Refinement Based on Explicit-Interpolation}
\label{subsubsection:explicit_refine}

The goal of our interpolation-based refinement for explicit-value analysis
is to determine a localized precision that is strong enough
to eliminate an infeasible error path in future explorations.
This criterion is fulfilled by the property of interpolants.
A second goal is to have a precision that is as weak as possible,
by creating interpolants that have a definition range as small as possible,
in order to be parsimonious in tracking variables and creating abstract states. 

\begin{algorithm}[t]
\begin{small}
\caption{$\mathsf{Refine}(\path)$}
\label{algorithm:extractPrecision}
\begin{algorithmic} 
\INPUT{infeasible error path $\path = \seq{(\op_1, \pc_1), ..., (\op_n, \pc_n)}$}
\OUTPUT{precision $\Pi$}
\VARDECL{interpolating constraint sequence $\cseqintpol$}

\STATE{$\cseqintpol := \seq{}$};
\STATE{$\Pi(\pc) := \emptyset$, for all program locations $\pc$};
\FOR{$i := 1$ to $n-1$}
  \STATE{$\cseq^+ := \seq{\op_{i+1}, ..., \op_n}$}
  \STATE{// inductive interpolation}
  \STATE{$\cseqintpol := \mathsf{Interpolate}(\cseqintpol \land \op_i, \cseq^+)$}
  \STATE{// extract variables from variable assignment that results from $\cseqintpol$}
  \STATE{$\Pi(\pc_i) := \big\{ x \big| (x, z) \in \SP{\cseqintpol}{\emptyset}$ 
          and $\bot \not= z \not= \top\big\}$}
\ENDFOR

\STATE{\bf return $\Pi$}
\end{algorithmic}
\end{small}
\end{algorithm}

\begin{figure*}[t]
\centering
\includegraphics[scale=0.5]{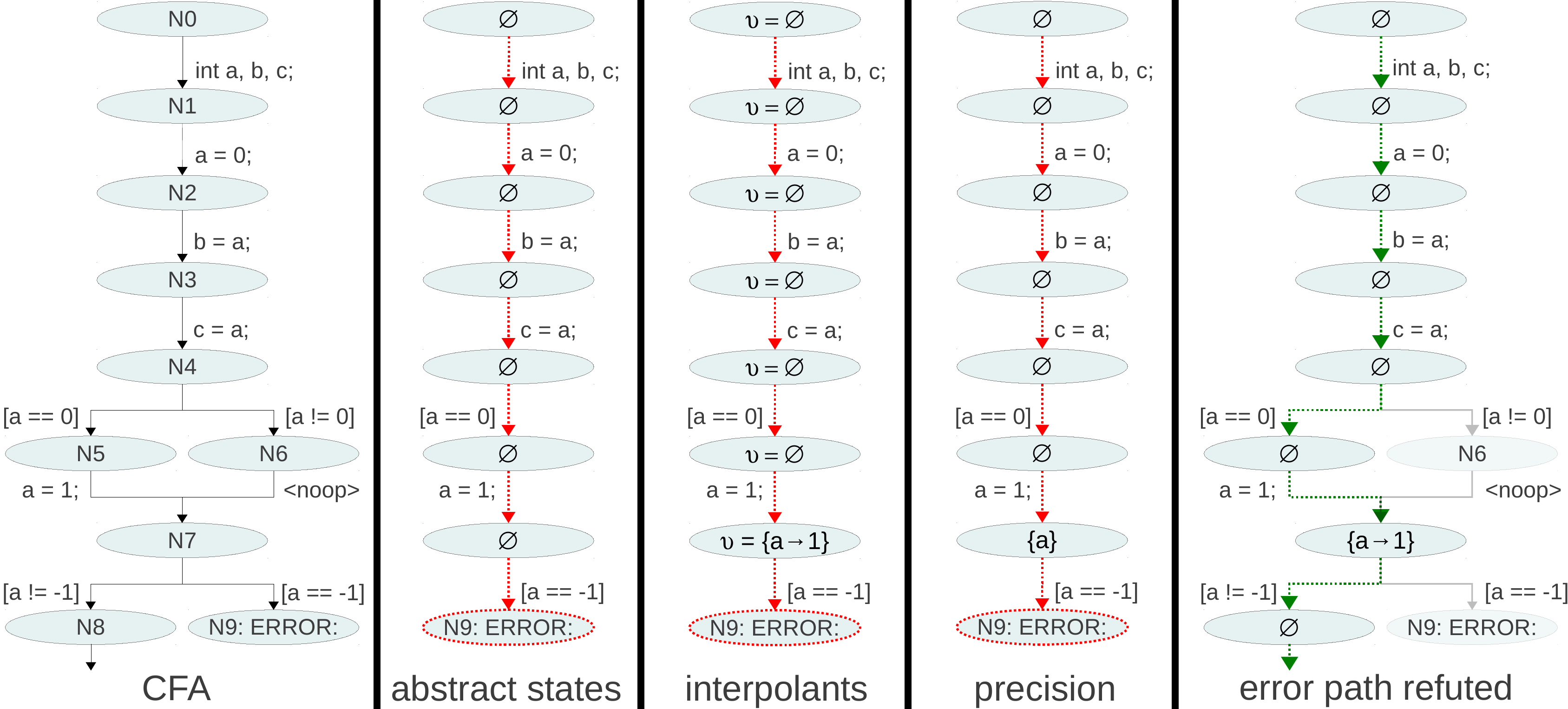}
\caption{
Illustration of one refinement iteration; from left to right: 
a simple example CFA, 
an infeasible error path with the abstract states annotated in the nodes (precision was empty, nothing is tracked),
the interpolated variable assignments annotated in the nodes, 
the precisions extracted from the interpolants annotated in the nodes,
and finally the CFA with the abstract states annotated in the nodes according to the new precision
(unreached nodes ---including error--- shown in gray)
}
\label{figure:explicit_interpolation}
\vspace{-4mm}
\end{figure*}

We apply the idea of interpolation for constraint sequences 
to assemble a precision-extraction algorithm:
Algorithm~$\mathsf{Refine}$ (Alg.~\ref{algorithm:extractPrecision}) 
takes as input an infeasible program path, 
and returns a precision for a program.
A further requirement is that the procedure computes \emph{inductive} interpolants~\cite{BLAST},
i.e., each interpolant along the path contains just enough information to prove 
the remaining path infeasible.
This is needed in order to ensure that the interpolants at the different locations achieve the goal
of providing a precision that eliminates the infeasible error path from further explorations.
For every program location~$\pc_i$ along an infeasible error path~$\path$, starting at $\pc_0$,
we split the constraint sequence of the path into a constraint prefix~$\cseq^-$,
which consists of the constraints from the start location~$l_0$ to~$\pc_i$, 
and a constraint suffix~$\cseq^+$, which consists of the path
from the location~$\pc_i$ to~$\pct$.
For computing inductive interpolants,
we replace the constraint prefix by the conjunction of the last interpolant and 
the current constraint.
The precision is extracted by computing the abstract variable assignment for 
the interpolating constraint sequence and assigning the relevant variables as precision
for the current location~$\pc_i$,
i.e., the set of all variables that are necessary to be tracked in order to 
eliminate the error path from future exploration of the state space.
This algorithm for precision extraction yields a parsimonious precision,
i.e., a precision containing \emph{just} enough information to exclude the infeasible error path,
and can be directly plugged-in as refinement routine of the \cegar algorithm (cf. Alg.~\ref{algorithm:cegar}).
Note that the repetitive interpolations are not an efficiency bottleneck.
The path is always finite, without any loops or branching, 
and thus, even a full-precision check can be decided efficiently.
Figure~\ref{figure:explicit_interpolation} illustrates the interpolation process 
on a simple example.

\subsection{Optimizations}
In our implementation, we added several optimizations to improve the performance of our approach.

\smallsec{ARG Pruning instead of Restart}
Our refinement routine~$\mathsf{Refine}$ (cf. Alg.~\ref{algorithm:extractPrecision}) 
returns a set of variables (precision)
that are important for deciding the reachability of the error location.
One of the ideas of lazy abstraction refinement~\cite{LazyAbstraction} 
is that the precision is only refined where necessary, i.e.,
only at the locations along the path that was considered in the refinement;
the other parts of the state space are not refined.
As mentioned in the discussion of the CEGAR algorithm (cf. Alg.~\ref{algorithm:cegar}),
it is not necessary to restart the exploration of the state space from scratch 
after a refinement.
Instead, we identify the descendant closest to the root of the abstract reachability graph (ARG)
in which the precision was refined,
and the re-exploration of the state space continues from there.
In total, this significantly reduces the number of tracked variables
per abstract state, which in turn leads to a more efficient analysis,
because it drastically increases the chance that a new abstract state is covered by an existing
abstract state.

\smallsec{Scoped Precision Refinement}
The precision for a program assigns to each program location the set of variables
that need to be tracked at that location, and
the interpolation-based refinement adds new variables 
precisely at the locations for which they were discovered during refinement.
In our experience, the number of refinements is reduced significantly
if we add a variable to the precision not only at the particular location for which it was discovered,
but at all locations in the local scope of the variable.
This helps to avoid adding a variable twice that can occur on two different branches.
By adding the variable to the precision ``in advance'' in the local scope,
we abbreviate some refinement iterations.
For example, consider Fig.~\ref{figure:explicit_interpolation} again.
After the illustrated refinement, another refinement step would be necessary,
in order to discover that variable~$\mathsf{a}$ needs to be tracked at location~$\mathsf{N4}$ as well
(to prevent the analysis from going through location~$\mathsf{N6}$).
By adding variable~$\mathsf{a}$ to the precision of all locations in the scope of variable~$\mathsf{a}$
immediately after the first refinement,
the program can be proved safe without further refinement.
This effect was also observed, and used, in the software model checker \blast~\cite{BLAST}.

\smallsec{Precise Counterexample Check}
In order to further increase the precision of our analysis,
we double-check all feasible error paths using bit-precise
bounded model checking (BMC)\,\footnote{In our implementation, we use CBMC~\cite{CBMC} as bounded model checker.},
by generating a path program~\cite{PathPrograms} for the error path and let the BMC confirm the bug.
Since the generated path program does not contain any loop or branching,
it can be verified efficiently.
If both our analysis and the bit-precise BMC report $\mathit{unsafe}$, then we report a bug.
If the BMC cannot confirm the bug, our analysis continues trying to find another error path.
This additional feature is available as a command-line option in our implementation.

\smallsec{Auxiliary Predicate Analysis}\label{smallsec:aux_pred_analysis}
As an additional option for further improvement of the analysis,
we implemented the combination with a predicate analysis, as outlined in existing work~\cite{CPAplus}.
In this combination, if the explicit-value analysis finds an error path,
this path is first checked for satisfiability in the predicate domain.
If the satisfiability check is positive,
the result $\mathit{unsafe}$ can be reported and the error path is returned;
if negative, then the explicit-value domain is not expressive enough to analyze that program path
(e.g., due to inequalities).
In this case, we ask the predicate analysis to refine its abstraction along that path,
which yields a refined predicate precision that eliminates the error path
but considering the facts along that path in the (more precise, and more expensive) predicate domain.
We need to parsimoniously use this feature
because the post-operations of the predicate analysis are
much more expensive than the post-operations of the explicit-value analysis.
In general, after a refinement step, either the explicit-value precision is refined (preferred)
or the predicate precision is refined (only if explicit does not succeed).

Using the concept of dynamic precision adjustment~\cite{CPAplus},
we also switch off the tracking of variables in the explicit-value domain
if the number of different values on a path exceeds a certain threshold.
After this, the predicate analysis will get switched on (by the above-mentioned mechanism)
and the facts on that path are further tracked using predicates.
This is important if the explicit-value analysis tries to unwind loops;
the symbolic, predicate-based analysis can often store a large number of values more efficiently.

Note that this refinement-based, parallel composition with precision adjustment
of the explicit-value analysis and the predicate analysis
is more powerful than a mere parallel product of the two analyses, 
because after each refinement, the explicit part of the analysis tracks exactly
what it is capable of tracking, while the auxiliary predicate analysis takes care of 
only those facts that are
beyond the capabilities of the explicit domain, resulting in a lightweight analysis on both ends.
Such a combination is easy to achieve in our implementation, because we use the
framework of configurable program analysis (CPA), which
lets the user freely configure such combinations.

\section{Experiments}
\label{section:experiments}

In order to demonstrate that our approach yields a significant practical
improvement of verification efficiency and effectiveness,
we implemented our algorithms and
compared our new techniques to existing tools for software verification.
In the following, we show that the application of abstraction, \cegar, and interpolation
to the explicit-value domain considerably improves the number of solved instances 
and the run time.
Combinations of the new explicit-value analysis with a predicate-based analysis
can further increase the number of solved instances.
All our experiments were performed on hardware identical to that of 
the SV-COMP'12~\cite{SVCOMP12},
such that our results are comparable to all the results obtained there.

\smallsec{Compared Verification Approaches}
For presentation, we restrict the comparison of our new approach to 
the SV-COMP'12 participants \blast, \satabs, and the competition winner \cpamemo,
all of which are based on predicate abstraction and CEGAR.
Furthermore, to investigate performance differences in the same tool environment,
we also compare with different configurations of \cpachecker.
%
The model checker \blast is based on predicate abstraction, and
uses a CEGAR loop for abstraction refinement.
The predicates for the precision are learned from counterexample paths
using interpolation.
The central data structure of the algorithm is an
ARG, which is lazily constructed and refined.
\blast won the category ``DeviceDrivers64'' in 
the SV-COMP'12, and got bronze in another category.
%
The model checker \satabs is also based on predicate abstraction and CEGAR,
but in contrast to \blast, it constructs and checks in every iteration of the CEGAR loop
a new boolean program based on the current precision of the predicate abstraction,
and does not use lazy abstraction or interpolation.
\satabs got silver in the categories ``SystemC'' and ``Concurrency'',
and bronze in another category.
The model checker \cpamemo is based on predicate abstraction, CEGAR, and interpolation,
but extends it with the concepts of adjustable-block encoding~\cite{ABE} and
block-abstraction memoization~\cite{CPACHECKERMEMO-COMP}.
\cpamemo won the category ``Overall'', got silver in two more categories, and bronze in another category.

We implemented our concepts as extensions of \cpachecker~\cite{CPACHECKER},
a software-verification framework based on configurable program analysis (CPA).
We compare with the existing explicit-value analysis 
(without abstraction, CEGAR, and interpolation)
and with the existing predicate analysis that is based on
boolean predicate abstraction, CEGAR, interpolation, and 
adjustable-block encoding~\cite{ABE}.
We used the trunk version of \cpachecker%
\footnote{\href{http://cpachecker.sosy-lab.org/}{http://cpachecker.sosy-lab.org}}
in revision 6615.

\smallsec{Verification Tasks}
For the evaluation of our approach,
we use all SV-COMP'12\,%
\footnote{\href{http://sv-comp.sosy-lab.org/2012/}{http://sv-comp.sosy-lab.org/2012}}
verification tasks that do not involve concurrency properties
(all categories except category ``Concurrency'').
All obtained experimental data as well as the tool implementation are available at
\href{http://www.sosy-lab.org/~dbeyer/cpa-explicit/}{{\small\tt http://www.sosy-lab.org/$\sim$dbeyer/cpa-explicit}}.

\smallsec{Quality Measures}
We compare the verification results of all verification approaches
based on three measures for verification quality:
First, we take the run time, in seconds, of the verification runs to measure the \emph{efficiency} of an approach.
Obviously, the lower the run time, the better the tool.
Second, we use the number of correctly solved instances of verification tasks
to measure the \emph{effectiveness} of an approach.
The more instances a tool can solve, the more powerful the analysis is.
Third, and most importantly,
we use the scoring schema of the SV-COMP'12
as indicator for the quality of an approach.
The scoring schema implements a community-agreed weighting schema,
namely, that it is more difficult to prove a program correct compared to finding a bug
and that a wrong answer should be penalized with double the scores that a correct answer would have achieved. 
For a full discussion of the official rules and benchmarks of the SV-COMP'12, 
we refer to the competition report~\cite{SVCOMP12}.
Besides the data tables, 
we use plots of quantile functions~\cite{SVCOMP12} for visualizing the number of solved instances and the verification time.
The quantile function for one approach
contains all pairs~$(x, y)$ such that the maximum run time of the 
$x$~fastest results is~$y$.
We use a~logarithmic scale for the time range from 1\,s to 1000\,s
and a linear scale for the time range between 0\,s and 1\,s.
In addition, we decorate the graphs with symbols at every fifth data point 
in order to make the graphs distinguishable on gray-scale prints.

\begin{table*}[t]
  \centering
  \begin{scriptsize}
  \begin{tabular}{l r r r r r r}
    \hline Category & \multicolumn{3}{c}{\cpaexp} & \multicolumn{3}{c}{~\cpaexpintpoltable} \\ \hline
 	& points & solved & time & ~points & solved & time	\\ \hline
ControlFlowInt	& 124 & 81 & 8400	& 123 & 79 & 780	\\ \hline
DeviceDrivers	& 53 & 37 & 63	& 53 & 37 & 69	\\ \hline
DeviceDrivers64	& 5 & 5 & 660	& 33 & 19 & 200	\\ \hline
HeapManipul	& 1 & 3 & 5.5	& 1 & 3 & 5.8	\\ \hline
SystemC	& 34 & 26 & 1600	& 34 & 26 & 1500 \\ \hline  
Overall	\phantom{\large 8}& 217 & 152 & 11000	& \normalsize{\textbf{244}} & \normalsize{\textbf{164}} & \normalsize{\textbf{2500}} \\
\hline
  \end{tabular}
 \end{scriptsize}
  \caption{Comparison with purely explicit, non-CEGAR approach}
  \label{tab:exp1}
\vspace{-4mm}
\end{table*}

\begin{figure}[t]
 \centering
 \includegraphics[width=0.4\textwidth]{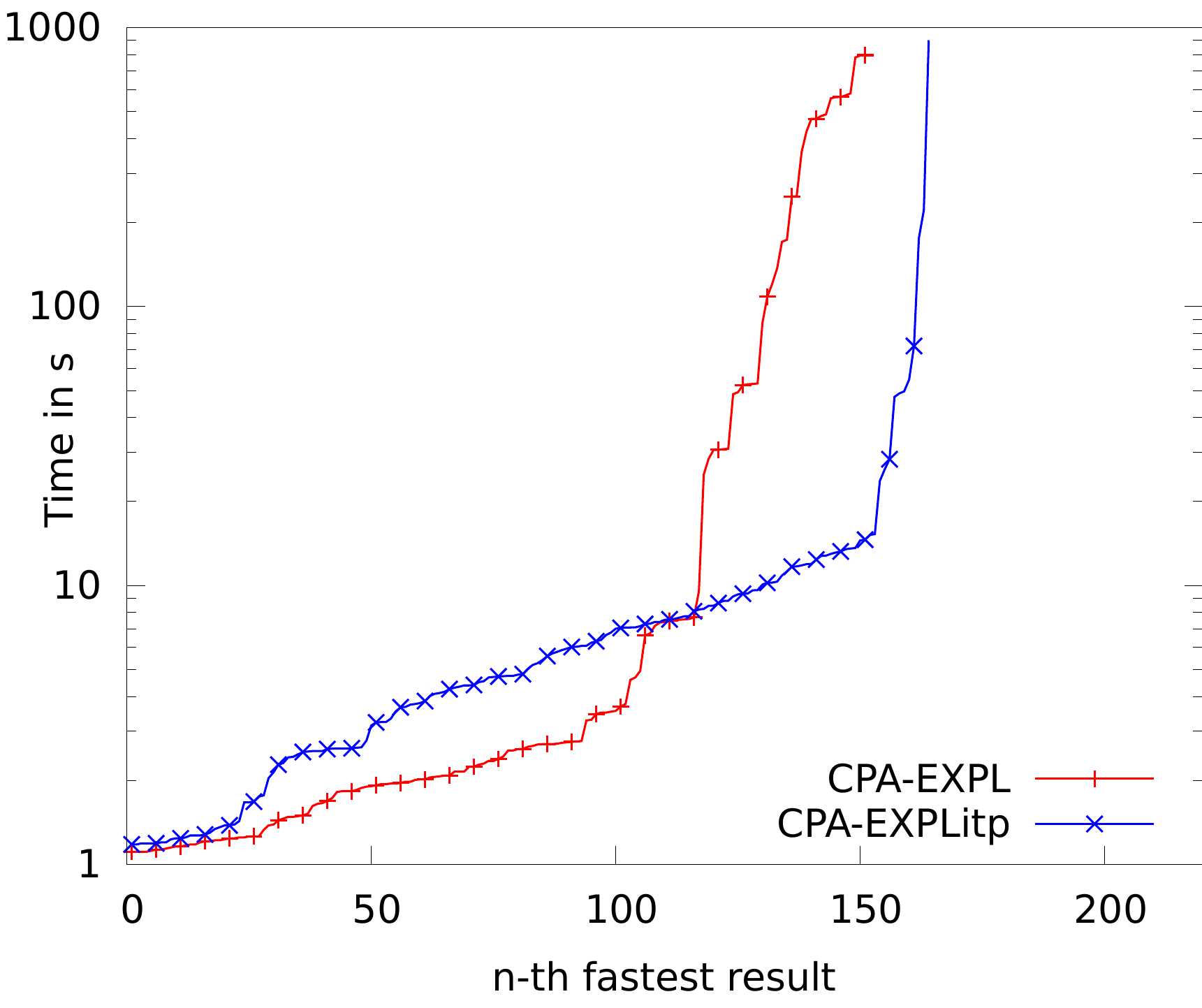}
\vspace{-2mm}
 \caption{Quantile plot: purely explicit analyses}
 \label{fig:exp1}
\vspace{-6mm}
\end{figure}

\smallsec{Improvements of Explicit-Value Analysis}
In the first evaluation, we compare two different configurations of the
explicit-value analysis:
\cpaexp refers to the existing implementation of a standard explicit-value analysis without abstraction and refinement, and
\cpaexpintpol refers to the new approach, which implements abstraction, CEGAR, and interpolation.
Table~\ref{tab:exp1} and Fig.~\ref{fig:exp1} show that the new approach
uses less time, solves more instances, and obtains more points in the SV-COMP'12 scoring schema.

\smallsec{Improvements of Combination with Predicate Analysis}
In the second evaluation, we compare the refinement-based explicit analysis
against a standard predicate analysis, as well as to the predicate analysis combined with \cpaexp and \cpaexpintpol, respectively:
\cpaabelf refers to a standard predicate analysis that \cpachecker offers (ABE-lf, \cite{ABE}), 
\cpaexpintpol refers again to the explicit-value analysis, which implements abstraction, CEGAR, and interpolation,
\cpaabelfexp refers to the combination of predicate analysis and explicit-value analysis without refinement,
and 
\cpaexpintpolabelf refers to the combination of predicate analysis
and explicit-value analysis with refinement.

\begin{figure}[t!]
\centering
\includegraphics[width=0.4\textwidth]{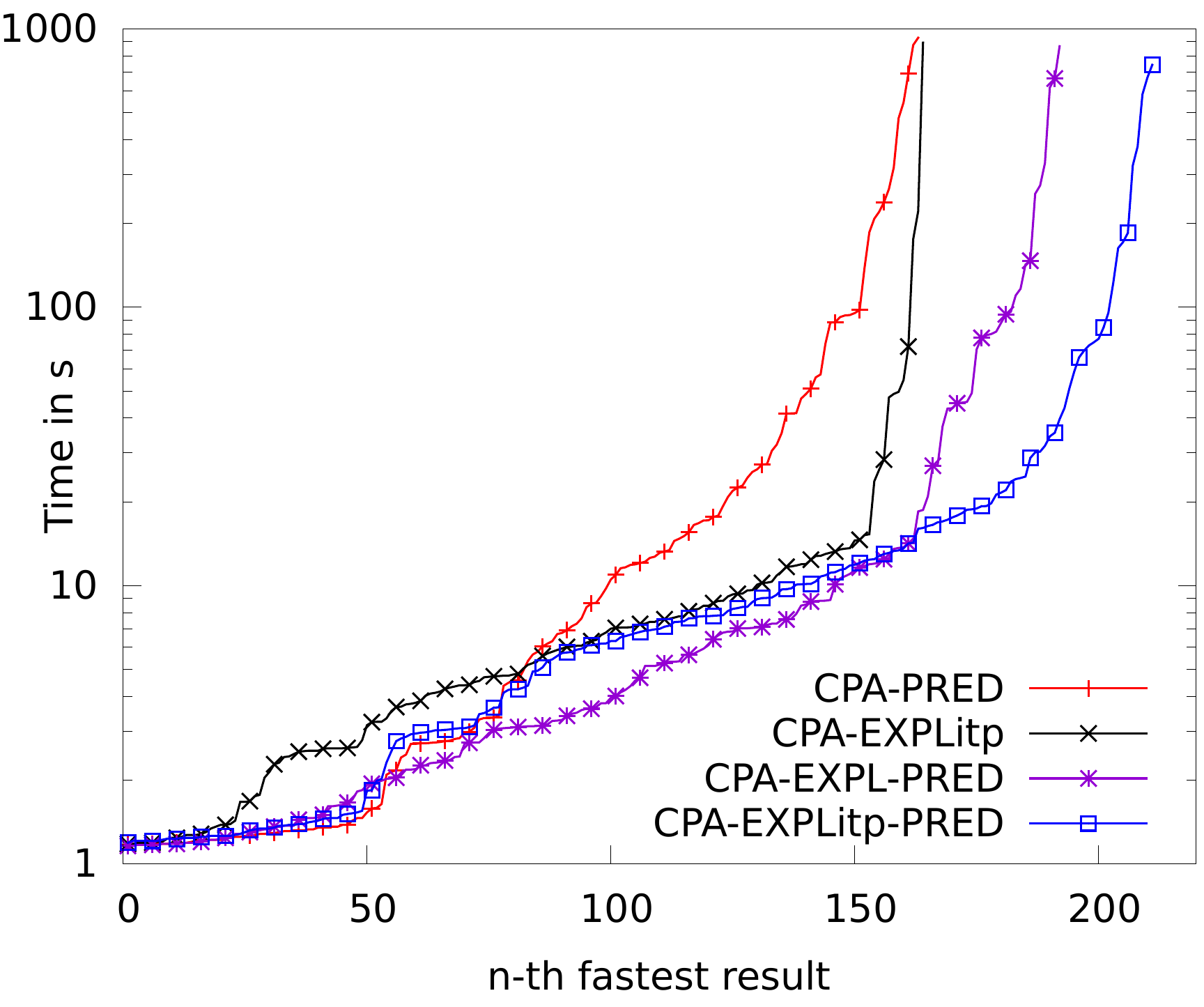}
\vspace{-2mm}
\caption{Quantile plot: comparison with predicate-based configurations}
\label{fig:exp2}
\vspace{-6mm}
\end{figure}

Table~\ref{table:table_CPA_EXPLitp_PRED} and Fig.~\ref{fig:exp2} show that the new combination approach
outperforms the existing approaches \cpaabelf and \cpaexpintpol in terms of solved instances and score.
The comparison with column \cpaabelfexp is interesting because it shows that the combination
of two analyses is an improvement even without refinement in the explicit-value analysis,
but switching on the refinement in both domains makes the new combination significantly more effective.

\begin{figure}[t!]
\vspace{3mm}
\centering
\includegraphics[width=0.4\textwidth]{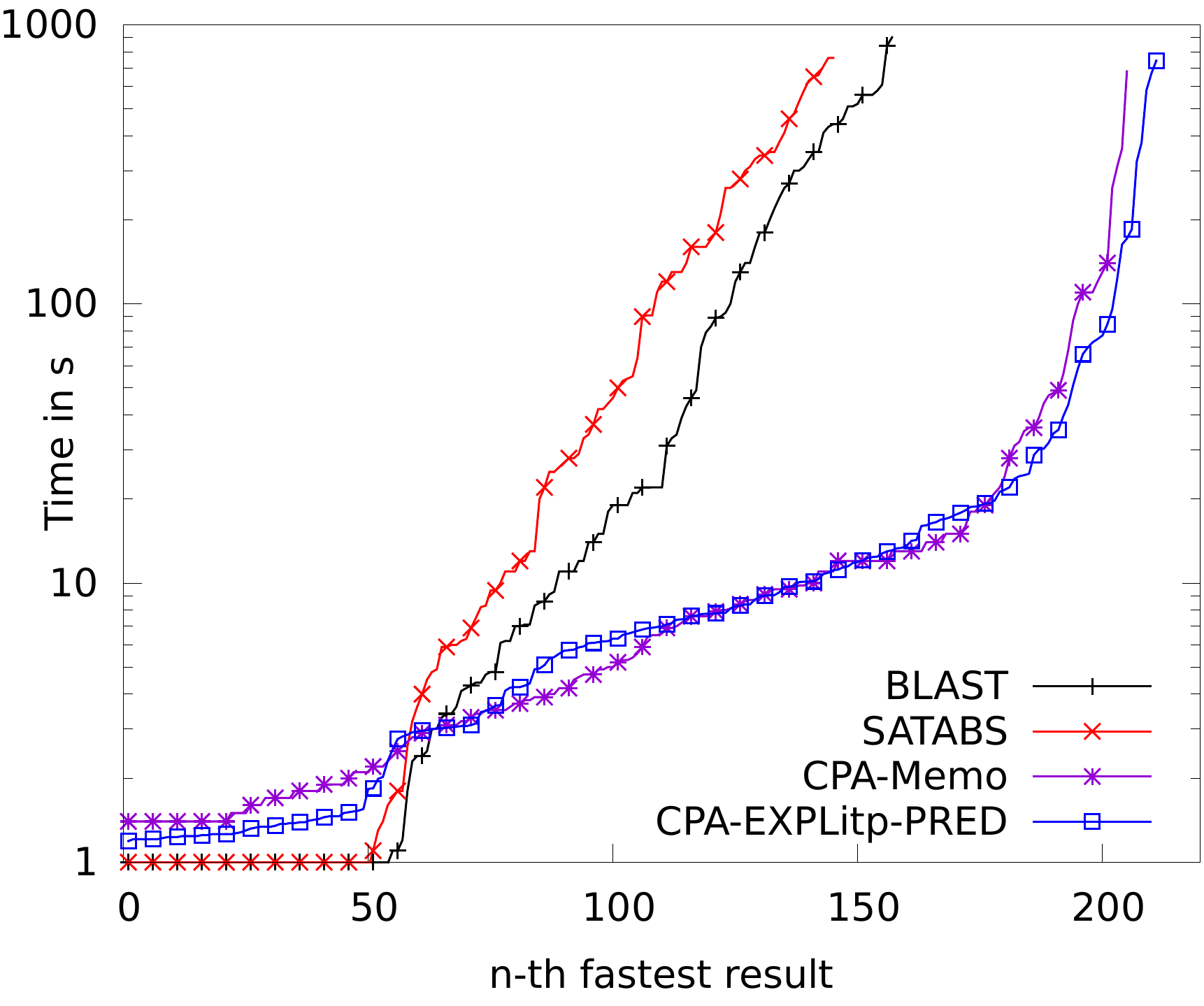}
\vspace{-2mm}
\caption{Quantile plot: comparison with three existing tools}
\label{fig:exp3}
\vspace{-6mm}
\end{figure}

\smallsec{Comparison with State-of-the-Art Verifiers}
In the third evaluation, we compare our new combination approach with three
established tools:
\blast refers to the standard \blast configuration that participated in the SV-COMP'12,
\satabs also refers to the respective standard configuration,
\cpamemo refers to a special predicate abstraction that is based on block-abstraction memoization, and
\cpaexpintpolabelf refers to our novel approach, which combines a predicate analysis (\cpaabelf)
with the new explicit-value analysis that is based on abstraction, CEGAR, and interpolation (\cpaexpintpol).
Table~\ref{table:table_overall} and Fig.~\ref{fig:exp3} show that the new approach outperforms
\blast and \satabs by consuming considerably less verification time, more solved instances, and a better score.
Even compared to the SV-COMP'12 winner, \cpamemo, our new approach scores higher.
It is interesting to observe that the difference in scores is much higher than the difference in 
solved instances: this means \cpamemo had many incorrect verification results,
which in turn shows that our new combination is significantly more precise.

\begin{table*} [t!]%
\newcommand\smallerfontsize\tiny%
\centering\scriptsize
\begin{tabular}{l r r r r r r r r r r r r}%
\hline Category & \multicolumn{3}{c}{\cpaabelf} & \multicolumn{3}{c}{~~\cpaexpintpoltable} & \multicolumn{3}{c}{~~\cpaabelfexp} & \multicolumn{3}{c}{~~\cpaexpintpolabelftable}\\\hline%
                                 & score & solved & time    & ~~score & solved & time   & ~~score & solved & time & ~~score                & solved                 & time 	\\ \hline
ControlFlowInt	                 & 103   &  70    & 2500	& 123     &  79    &  780	& 131     & 85     & 2600 & 141                    & 91                     & 830		\\ \hline
DeviceDrivers	                 &  71   &  46    &   80	&  53     &  37    &   69	&  71     & 46     &   82 &  71                    & 46                     &  87		\\ \hline
DeviceDrivers64	                 &  33   &  24    & 2700	&  33     &  19    &  200	&  10     & 11     & 1100 &  37                    & 24                     & 980		\\ \hline
HeapManipul	                     &   8   &   6    &   12	&   1     &   3    &    5.8	&   6     &  5     &   11 &   8                    &  6                     &  12		\\ \hline
SystemC	                         &  22   &  17    & 1900	&  34     &  26    & 1500	&  62     & 45     & 1500 &  61                    & 44                     & 3700		\\ \hline
Overall	\phantom{{\large 8}}   & 237   & 163    & 7100	    & 244     & 164    & 2500	& 280     & 192    & 5300 & {\normalsize\bf 318} & {\normalsize\bf 211} & 5600	
\\\hline%
\end{tabular}%
\caption[Comparison with predicate-based configurations]{Comparison with predicate-based configurations}%
\label{table:table_CPA_EXPLitp_PRED}%
\end{table*}%

\begin{table*} [t!]%
\newcommand\smallerfontsize\tiny%
\centering\scriptsize
\begin{tabular}{l r r r r r r r r r r r r}%
\hline Category & \multicolumn{3}{c}{\blasttable} & \multicolumn{3}{c}{\satabstable} & \multicolumn{3}{c}{\cpamemo} & \multicolumn{3}{c}{~~\cpaexpintpolabelftable}\\\hline%
                             & score & solved & time   & ~~score & solved & time  & ~~score & solved & time & ~~score                     &                       solved & time	\\ \hline
ControlFlowInt               & 71    & 51     & 9900   & 75      & 47     & 5400  & 140     & 91     & 3200	&                         141 &                           91 & 830	\\ \hline
DeviceDrivers	             & 72    & 51     & 30	   & 71      & 43     & 140	  & 51      & 46     & 93	&                          71 &                           46 & 87	\\ \hline
DeviceDrivers64	             & 55    & 33     & 1400   & 32      & 17     & 3200  & 49      & 33     & 500	&                          37 &                           24 & 980	\\ \hline
HeapManipul	                 & --    & --     & --	   & --      & --     & --	  & 4       & 9      & 16	&                           8 &                            6 & 12	\\ \hline
SystemC	                     & 33    & 23     & 4000   & 57      & 40     & 5000  & 36      & 30     & 450	&                          61 &                           44 & 3700	\\ \hline
Overall	\phantom{\large 8} & 231   & 158    & 15000  & 235     & 147    & 14000 & 280     & 209    & 4300	& \normalsize{\textbf{318}} & \normalsize{\textbf{211}} & 5600	

\\\hline%
\end{tabular}%
\caption[Comparison with three existing tools]{Comparison with three existing tools}%
\label{table:table_overall}%
\end{table*}%

\vspace{-1mm}
\section{Conclusion}
\vspace{-1mm}

The surprising insight of this work is that it is possible to achieve
---without using sophisticated SMT-solvers during the abstraction refinement---
a performance and precision that can compete with the world's leading
symbolic model checkers, which are based on SMT-based predicate abstraction.
We achieved this by incorporating the ideas of abstraction, 
counterexample-guided abstraction refinement, 
lazy abstraction refinement, and interpolation into
a standard, simple explicit-value analysis.

We further improved the performance and precision by
combining our refinement-based explicit-value analysis with a predicate analysis,
in order to benefit from the complementary advantages of the methods.
The combination analysis dynamically adjusts the precision~\cite{CPAplus}
for an optimal trade-off between the precision of the explicit analysis
and the precision of the auxiliary predicate analysis.
This combination out-performs state-of-the-art model checkers,
witnessed by a thorough comparison on a standardized set of benchmarks.

Despite the overall success of our new approach, 
individual instances of benchmarks show different
performance with different configurations --- i.e., either with or without \cegar.
Therefore, a general heuristic for finding a suitable strategy for a single
verification task would be beneficial.
Also, we envision better support for pointers and data structures,
because our interpolation approach can be efficiently applied even with high precision.
Moreover, we so far only combined our interpolation approach with an auxiliary predicate analysis in the ABE-lf configuration,
and we have not yet tried to combine this with the superior block-abstraction memoization (ABM) \cite{CPACHECKERMEMO-COMP} technique.
Finally, we plan to extend our interpolation approach to other abstract domains
like intervals.
\balance


\end{document}